# The role of double ionization on the generation of doubly charged ions in copper vacuum arcs: insight from particle-in-cell/direct simulation Monte Carlo methods

Wei Yang[a)], Qianhong Zhou, Qiang Sun, Wenyuan Yang, and Zhiwei Dong

Institute of Applied Physics and Computational Mathematics, Beijing, China,100094
[a)]Email:yangwei861212@126.com

Metal vapour vacuum arcs are capable to generate multiply charged metallic ions, which are widely used in fields such as ion deposition, ion thrusters, and ion sources, etc. According to the stationary model of cathode spot, those ions are generated by electron-impact single ionization in a step-wise manner, which is M -> $M^+$ -> $M^{2+}$ -> ... mainly. This paper is designed to study quantitatively the role of double ionization M -> $M^{2+}$ in the breakdown initiation of copper vacuum arcs. A direct simulation Monte Carlo (DSMC) scheme of double ionization is proposed and incorporated into a 2D particle-in-cell (PIC) method. The super-particles of $Cu^{2+}$ ions generated from different channels are labelled independently in the PIC-DSMC modelling of vacuum arc breakdown. The cathode erosion rate based on PIC modelling is about 40μg/C in arc burning regime, which agrees well with previous experiments. The temporal discharge behaviours such as arc current, arc voltage, and ionization degree of arc plasma, are influenced with or without double ionization negligibly. However, additional $Cu^{2+}$ ions are generated near the cathode in breakdown initiation from the double ionization channel, with a lower kinetic energy on average. Therefore, the results on spatial distribution and energy spectra of $Cu^{2+}$ ions are different with or without double ionization. This paper provides a quantitative research method to evaluate the role of multiply ionization in vacuum arcs.

## 1. Introduction

Metal vapour vacuum arcs are widely used in deposition systems [1], circuit breakers [2], ion sources [3], electrical thrusters [4], and etc. The vacuum arc belongs to arc discharge regimes, which manifests itself as a high current and low voltage discharge in current-voltage characteristic [5]. Although the burning voltage of the arc is basically several tens of volts, the initiation of the vacuum breakdown requires voltage up to several thousands of volts. The reason is that surface electric field on the order of $10^9$ V/m is anticipated for the explosive emission or thermo-field emission of electrons from the cathode [6]. Accompanying the emission of electrons, the metal vapours/plasmas are emitted from the cathode surface. The essential part of vacuum arc is cathode spot, which consists of highly ionized, multiply charged metallic plasma. In fact, metallic ions with average charge states up to +6 were observed in the short-pulse high-current discharges in vacuum [7, 8].

The generation mechanism of such highly charged ions was not clear, partly due to the controversial models of cathode spots. One is a stationary model [9], in which the ions are mainly generated in a mild manner by electron-impact ionization of metallic vapours. The other is a non-stationary model, in which the ions are mainly generated in a wild manner by explosive emission of dense plasma [10, 11]. The two models may be both present in different stages of the life cycle of cathode spots, depending on the current densities, the local heating, the surface morphology, and other factors. For example, in the non-stationary model, the dense plasma may result from a rapid phase-transition of the cathode material directly from solid state to non-ideal plasma [12], and the local temperature is beyond the critical point of the cathode material. While in the stationary model, the metallic vapours are emitted from the already melted cathode surface, and the local temperature is in the range from melting point to boiling point of the cathode material [13-15].

This paper focuses on the stationary model, and examines the possibility of double ionization on the generation of $Cu^{2+}$ ions by particle-in-cell (PIC)/direct simulation Monte Carlo (DSMC) method. The PIC-DSMC method is a first-principle calculation tool, which is widely used in simulation study of vacuum arcs [16-20]. In the previous researches, the role of double ionization Cu -> $Cu^{2+}$ was presumably neglected without further justification from either experiment or modelling studies. Most studies took the step-wise ionization Cu -> $Cu^{+}$ -> $Cu^{2+}$ -> ... for granted in vacuum arc modelling. In this paper, the generation of doubly charged copper ions $Cu^{2+}$ directly from neutrals will be studied for at least two reasons. The first reason is that electrons can gain enough energy in the vacuum breakdown stage to overcome the double ionization threshold when they collide upon the neutrals. The second is that in the generation of $Cu^{2+}$, the double ionization is a single step process, while the step-wise ionization is a two-step consecutive process. The study here is expected to give firm evidence as what role the double ionization plays in the generation of multiply charged ions.

## 2. Numerical model

### 2.1 General description of the PIC-DSMC method

Cathode plasma is composed of large amounts of particles with different species i.e. electrons, multiply charged ions and neutrals. For each species, temporal evolution $t$ of distribution function $f(\vec{r},\vec{v},t)$ in real space $\vec{r}$ and velocity space $\vec{v}$ satisfies the nonlinear Boltzmann equation (BE) [21]

$$\frac{\partial f(\vec{r},\vec{v},t)}{\partial t} + \vec{v} \cdot \nabla f + \frac{d\vec{v}}{dt} \cdot \nabla_v f = \left(\frac{\partial f}{\partial t}\right)_{coll}. \tag{1}$$

Because the cathode plasma is highly ionized, the number densities of neutral particles may be very close or even less than that of charged particles, and cannot be viewed as background. The DSMC method is used to describe the collision operator in the right-hand-side of Eq. (1)

$$\left(\frac{\partial f}{\partial t}\right)_{coll} = \sum_q \iint (f'_p f'_q - f_p f_q) w_{pq} d\sigma_{pq} dv_q, \quad (2)$$

in which the distribution functions for all species need to be solved. This is in contrast to the test-particle MCC method for linear Boltzmann equations [22]

$$\left(\frac{\partial f}{\partial t}\right)_{coll} = \iint (f'_p G'_g - f_p G_g) w_{pg} d\sigma_{pg} dv_g, \quad (3)$$

in which the distribution function for neutrals $G_g$ is usually assumed to be Maxwell-like and is not solved. The collision processes are simulated by a Monte Carlo technique over all colliding super-particles inside each cell one by one. Taken binary collision as an example, a pair of super-particles is chosen randomly every time step, which are called projectile and target respectively. The mass of the projectile and target particle is denoted as $m_1$ and $m_2$, and the velocity is denoted as $v_1$ and $v_2$ respectively. The collision probability $p_{coll}$ is calculated as

$$p_{coll} = 1 - \exp(-\Delta t_{coll} |\vec{u}| n_t \sigma(\varepsilon_{rel})) \approx \Delta t_{coll} |\vec{u}| n_t \sigma(\varepsilon_{rel}), \quad (4)$$

In Eq. (4), $\Delta t_{coll}$ is the collision time step, $\vec{u} = \vec{v}_1 - \vec{v}_2$ is the relative velocity, $n_t$ is the number density of the target particle, $\varepsilon_{rel}$ is the collision energy of relative motion between projectile and target, and $\sigma$ is the integrated cross section. The collision occurs if the collision probability $p_{coll}$ is larger than a random number $R_{01}$ evenly distributed between [0, 1). By enforcing the momentum conservation law during the collision, the post-collision velocity can be written as

$$\vec{v}_1' = \vec{v}_1 + (\vec{u}' - \vec{u}) m_r / m_1,$$
$$\vec{v}_2' = \vec{v}_2 - (\vec{u}' - \vec{u}) m_r / m_2. \quad (5)$$

In Eq. (5), the superscript prime means velocity after collision, and $m_r = m_1 m_2 / (m_1 + m_2)$ is the reduced mass. By enforcing the energy conservation law during the collision, the magnitude of post-collision relative velocity $\vec{u}'$ is determined by

$$|\vec{u}'|^2 = |\vec{u}|^2 + 2\Delta E / m_r. \quad (6)$$

In Eq. (6), $\Delta E$ is the change of kinetic energy during the collision, and it equals to zero for elastic collision as a special case. The deflection of $\vec{u}$ is defined by the scattering and azimuth angles. The scattering angle $\chi$ is determined by the normalized differential cross section $I$ through another random number $R_{01}$. In all cases of electron-copper collision here, the isotropic scattering is assumed.

$$2\pi \int_0^\chi I(v,\chi')\sin\chi' \,d\chi' = R_{01} \tag{7}$$

The azimuth angle is set randomly distributed between 0 and $2\pi$.

Regarding for the left-hand-side of the BE, the motion of non-relativistic super-particle satisfies Newton's second law, and the details of the numerical scheme can be found in paper on PIC method [23]. The electric field inside the plasma is calculated by Poisson equation

$$\frac{1}{\varepsilon_0}\nabla^2\phi = -\rho. \tag{8}$$

In Eq. (8), $\varepsilon_0$ is the permittivity in vacuum and $\rho$ is charge density with contribution from all charged particles. The Poisson equation and BE are coupled through the electric field/potential and charge density with each other. The field is interpolated from mesh grids to particle position, and the charge density is scattered from particle position $\vec{r}_j$ to mesh grids $\vec{r}_k$. The assignment of charge density can be written as follows

$$\rho_k = \sum_\alpha q_\alpha W_\alpha \sum_j S(\vec{r}_k, \vec{r}_{j\alpha})/V_j. \tag{9}$$

In Eq. (9), $W_\alpha$ is the weight of particle $\alpha$, $S(\vec{r}_k, \vec{r}_j)$ is the shape factor, and $V_j$ is the cell volume. The popular strategy for shape factor is the so-called cloud-in-cell (CIC) scheme, which is a first order weighting scheme. In a 2D rectangle grid, the CIC scheme is written as

$$S(\vec{r}_k, \vec{r}_j) = \begin{cases} 0, |x_j - x_k| \geq \Delta x \text{ or } |y_j - y_k| \geq \Delta y \\ (1-|x_j - x_k|/\Delta x)(1-|y_j - y_k|/\Delta y), \text{otherwise} \end{cases}. \tag{10}$$

The cell volume is calculated by the method proposed by Verboncoeur [24]. At the same time, the interpolation of field from mesh grids to particle position is done in an inversely analogous way, which is consistent to that of charge density.

## 2.2 Detailed DSMC scheme of double ionization process

The DSMC scheme of double ionization for equally weighted super-particles is shown below. For not equally weighted particles, either the rejection method [25] or the merging method [26] can be consulted. To not lose generality, the process of double ionization is written as follows

$$e + A \Rightarrow e + A^{2+} + 2e. \tag{11}$$

The velocities of the two reactants are known, and the DSMC scheme is to find out the post-collision velocities for the four products. A strict formulation would require the complete knowledge of differential ionization cross sections to determine the energy partition and scattering angles between primary and secondary electrons. Unfortunately, those information are not available right now even for a single incident energy of electron impacting on copper. Therefore, the double ionization is decoupled into four-stage binary collisions, and the scheme for each stage is quite mature. In the first stage, the neutral A undergoes an inelastic collision, and the electron loses the energy of double-ionization threshold

$$e + A \Rightarrow e_1 + A^* \tag{12}$$

In the second stage, the meta-stable atom A* splits into an electron and a doubly charged ion, similar to auto-ionization. In this stage, the new born particles inherit the velocity and position of the old particle.

$$A^* \Rightarrow e_2 + A^{2+} \tag{13}$$

In the third stage, the primary $e_1$ and secondary $e_2$ undergo elastic collision, similar to Coulomb scattering

$$e_1 + e_2 \Rightarrow e_3 + e_4 \tag{14}$$

In the last stage, one of the electron splits into two electrons, and double ionization is completed.

$$e_3 \Rightarrow e_5 + e_6 \tag{15}$$

It's noted that the second stage and the last stage does not conserve electric charge and mass in a single stage. However, the total processes conserve charge, momentum, energy, and mass by the combination of four-stage collisions.

The splitting scheme in the last stage is shown as an example. To enforce the conservation of momentum and energy simultaneously, the velocity vectors after splitting form the two edges of a rectangle, and the velocity vector before collision is the diagonal line of the rectangle. The deflection of velocity between $e_5$ and $e_3$ is denoted as scattering $\chi$ and the azimuth $\psi$ angles. To find out the velocity of $e_5$, the following steps are used. The first step shortens the velocity vector $\vec{v}$ of $e_3$

$$\vec{v} \rightarrow \alpha \vec{v}, \alpha = \cos \chi \tag{16}$$

The second step transforms the velocity from laboratory frame to a local frame, where only z-component of the velocity vector is non-zero.

$$\alpha \vec{v} \rightarrow \ddot{R}(\theta,\varphi)(\alpha \vec{v}) = \alpha(0,0,|\vec{v}|) \tag{17}$$

In Eq. (17), $\theta$ is the angle between $\vec{v}_z$ and $\vec{v}$ before collision, and $\varphi$ is the angle between $\vec{v}_x$ and $\vec{v}_\perp = \vec{v}_x + \vec{v}_y$ before collision.

The third step scatters the velocity in local frame

$$\ddot{R}^{-1}(\chi,\psi)\ddot{R}(\theta,\varphi)(\alpha \vec{v}) \tag{18}$$

The last step transforms the velocity from the local frame back to the laboratory frame

$$\ddot{R}^{-1}(\theta,\varphi)\ddot{R}^{-1}(\chi,\psi)\ddot{R}(\theta,\varphi)(\alpha\vec{v}) \quad (19)$$

The transforming matrix can be written as

$$\ddot{R}(\theta,\varphi) = \begin{pmatrix} \cos\theta\cos\varphi & \cos\theta\sin\varphi & -\sin\theta \\ -\sin\varphi & \cos\varphi & 0 \\ \sin\theta\cos\varphi & \sin\theta\sin\varphi & \cos\theta \end{pmatrix} \quad (20)$$

## *2.3 Simulation models*

The double ionization is incorporated within 2D3V PIC-DSMC methods in cylindrical geometry developed previously [27]. The modelling starts from complete vacuum between two plane electrodes, and the domain is $z$=6 μm and $r$=24 μm. The spatial step is 50 nm both in the $z$ and $r$ direction. The electron time step is 1 fs, and time step of heavy particle is ten times that of electrons. The tracked super-particles include electrons, Cu neutrals, $Cu^+$, $Cu^{2+}$, $Cu^{3+}$ and $Cu^{4+}$, and all super-particles share the same weight of 100. Both simulation sets include electron-neutral elastic collision, Coulomb collision, charge exchange collision between neutral and single charged ion, and neutral-neutral elastic collision, besides ionization collision. In the first set of simulation, only step-wise ionization from Cu to $Cu^{4+}$ is considered: Cu -> $Cu^+$ -> $Cu^{2+}$ -> $Cu^{3+}$ -> $Cu^{4+}$. In the second set, the double ionization is added for the process of electron impact of copper neutral, and the $Cu^{2+}$ super-particles generated by single ionization $Cu^+$ -> $Cu^{2+}$ and double ionization Cu -> $Cu^{2+}$ channels are labelled independently. The cross sections for those import channels that are relevant to double ionization are from [28-30] and plotted in Figure 1. Other cross sections used in this simulation can be found in a separate paper [31] and references there in.

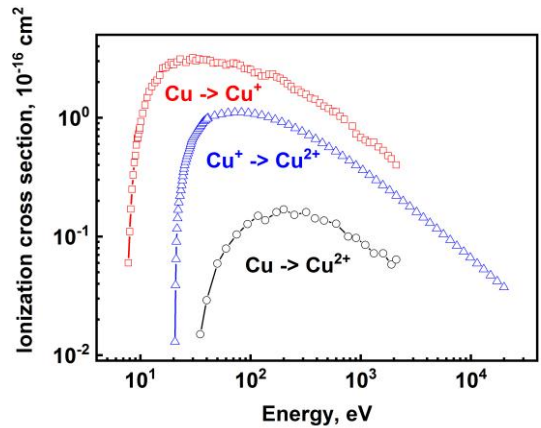

Figure 1. The cross sections [28-30] for those import channels that are relevant to double ionization.

A module of external circuit is coupled to the PIC simulation of plasmas. The cathode is grounded, and the anode is connected to a high voltage source 2.9 kV with external RC circuit with parameters: R= 1 kΩ and C=1 pF. To mimic the initiation of vacuum breakdown, an artificial tip with width 200 nm is located in the centre of the cathode with a nominal field enhancement factor of 35. In the initial stage, the tip-enhanced electric field is capable to extract electrons from the cathode surface, in which the electron current density is calculated by Fowler-Nordheim formalism. The neutral flux emitted from the cathode is set by a constant ratio of 3% to the emitted electron current density. No ions are injected in the surface or volume of the simulation domain, but are generated by electron-impact ionization of neutrals/ions. As ion densities build up, extra neutral fluxes are sputtered into the simulation domain by energetic ions incident upon the electrodes with sputtering yield given by Yamamura and Tawara [32].

The ratio of emitted neutral flux to electron current density 3% is quite close to that from molecular dynamics study on the erosion of a copper nanotip by arc plasma (2.5±0.3) % [33]. Although the nominal ratio 3% is set as a constant, the calculated result on specific cathode erosion rate varies with time between 20~40 μg/C, as shown in Figure 2 for the first simulation set. The erosion rate is calculated as the ratio between transferred mass and transferred charge. The reason for time-varying erosion rate is ascribed to the time-dependent returning flux of heavy particles and the sputtering flux. When the arc settles down to a steady burn at a time later than 0.6 ns, the cathode erosion rate reaches a stable value. The calculated value by PIC-DSMC (~40 μg/C) agrees well with the measured experimental data 35~40 μg/C from different groups [34, 35].

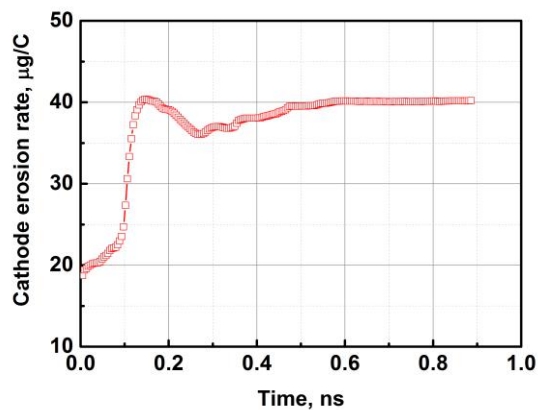

Figure 2. Specific cathode erosion rate based on the calculated results of the first simulation set.

# 3. Results and discussion

In this section, the simulation results from both sets are compared, in order to unveil the role of double ionization during the vacuum breakdown. Firstly, we would like to study the temporal evolution of macroscopic behavior such as current-voltage characteristic, ionization degree, and average charge state of arc discharge. Next, the microscopic behavior of spatial distribution of number density $Cu^{2+}$ ions at different stage of arc discharge is studied. Lastly, the energy spectra of $Cu^{2+}$ ions at different stage are also investigated. The super-particles of $Cu^{2+}$ ions generated by different ionization channels are labeled independently to trace their effects.

## 3.1 Temporal evolution of macroscopic discharge behaviours

Figure 3 compares the temporal evolution of (a) arc voltage *V*, and (b) arc current *I* of both simulation sets. It is not strange that for both macroscopic parameters *IV*, negligible difference is observed during the vacuum breakdown with or without consideration of double ionization. During the discharge, the dropping of the voltage between electrodes is calculated self-consistently with the external circuit. According to the simulation results, the current-voltage characteristics can be divided into three stages: the initiation, the breakdown, and the burn process. Because this paper focused on the generation of multiply charged ions, the extinction process of cathode spot is not simulated to save the computation time. In the initiation process, the gap voltage between electrodes is high and the current is very low, although local low-density plasma is already produced in the near-cathode region. In the breakdown process, the voltage drops and the current increases sharply, and the local plasma propagates toward the anode to form a conductive channel. In the burn process, a low arc voltage and a high arc current are reached and maintained in a steady state.

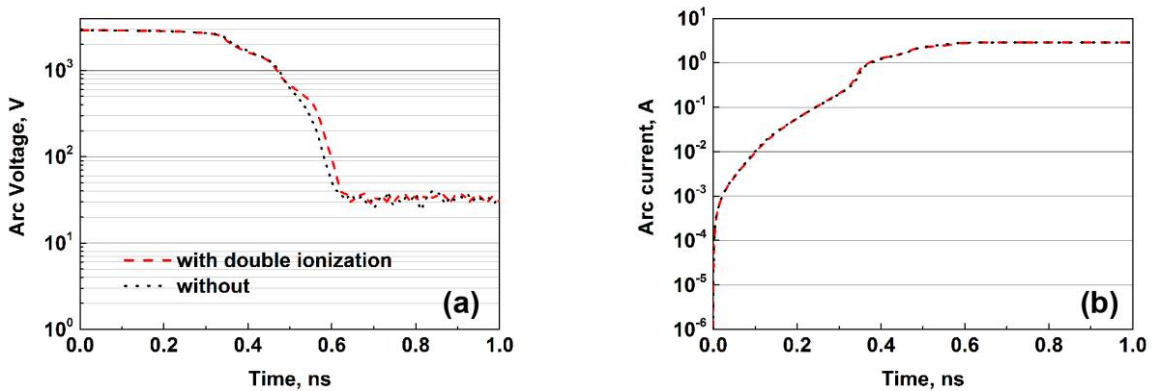

Figure 3. Temporal evolution of (a) arc voltage, and (b) arc current of both simulation sets.

Figure 4 compares the time evolution of (a) ionization degree $\gamma$, and (b) average charge state $\bar{Q}$ of ions for both simulation sets, which are calculated according to the particle number $N_q$ with different charge $q$ as follows

$$\gamma = \sum_{q=1}^{4} N_q / \sum_{q=0}^{4} N_q, \bar{Q} = \sum_{q=1}^{4} q N_q / \sum_{q=1}^{4} N_q. \tag{21}$$

The temporal behaviour of ionization degree and average ion charge state show similar trend with that of current-voltage characteristic: the influence of double ionization is negligible. In the initiation process, the ionization degree quickly increases, and the average charge state is about unity: the generated ions are dominated by single ionized $Cu^+$. In the breakdown process, the ionization degree decreases slightly, because the ionization of higher charged ions is much difficult than the ionization of lower charged ones, considering the cross sections of impact-ionization. At the same time, the average charge state increases to a maximum above 2.5 and then decreases. In the burn process, the ionization degree of the vacuum arc plasma reaches a quasi-steady state of about 60%, and the average charge state of ions is approximately 2.0. The temporal decrease of average state from about 2.5 to 2.0 agrees with the measured value from about 2.3 to 1.9 by Oks *et al.* at low current (Fig.6 in Ref. [36]).

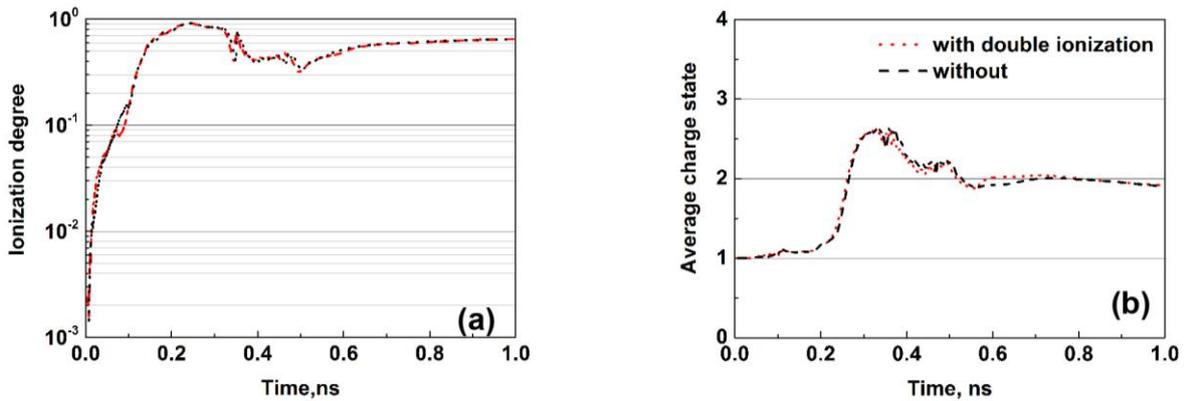

Figure 4. Temporal evolution of (a) ionization degree (in log-scale), and (b) average charge state of ions (in linear scale) of both simulation sets.

To explain the above results, the super-particle numbers of $Cu^+$ and labelled $Cu^{2+}$ based on the calculated results of the second simulation set are shown in Figure 5. The number of $Cu^+$ increases with time very quickly due to electron-impact ionization of Cu neutrals (black solid line). Once the number of super $Cu^+$ ions have aggregated to more than 100, the possibility of single ionization of $Cu^+$ to generate $Cu^{2+}$ is quite high (green dash line). However, the possibility of double ionization of Cu neutrals to generate $Cu^{2+}$ occurs slightly later (blue dotted line), because a large amount of Cu neutrals are destroyed in the single ionization process. The inset shows the ratio of $Cu^{2+}$ generated by double ionization $Cu \rightarrow Cu^{2+}$ to that generated by single ionization $Cu^+ \rightarrow Cu^{2+}$. Only in the early stage of arc initiation, the number of $Cu^{2+}$ generated by double ionization is comparable to that generated by single ionization. After 0.2 ns, the number of $Cu^{2+}$ generated by double ionization is less than 10% of that generated by single ionization.

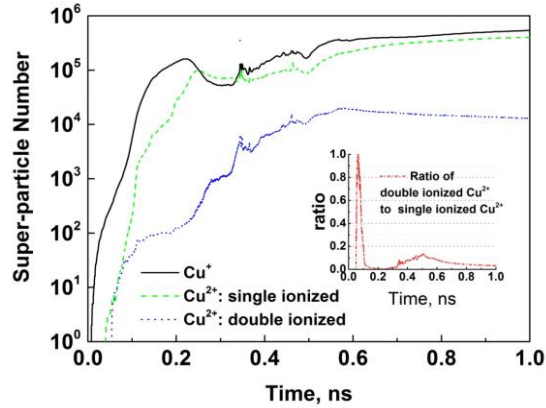

Figure 5. Super-particle number of $Cu^+$ and labelled $Cu^{2+}$ based on the calculated results of the second simulation set, and the inset shows the ratio of $Cu^{2+}$ generated by double ionization to that generated by single ionization.

## 3.2 Spatial distribution of $Cu^{2+}$ number density

Figure 6 compares the spatial distribution of $Cu^{2+}$ number density of both simulation sets at a time of (a) 0.3 ns and (b) 0.8 ns in the entire simulation domain, respectively. The two time points are representative of breakdown initiation and stable arc. The results at 0.3 ns show a slight difference without (upper part) and with (lower part, mirror image along the symmetric axis $r=0$ for comparison) double ionization taken into account. With double ionization, there are extra $Cu^{2+}$ ions located on the periphery of $Cu^{2+}$ cloud, as indicated by the red arrow in Figure 5(a) in the $z$ direction. In Figure 5(b) at 0.8 ns when the conductive channel is formed between the electrodes, no obvious difference can be found for the spatial distribution of $Cu^{2+}$ ions.

The difference at 0.3 ns is further investigated by examining the spatial distribution of number density of independently labelled $Cu^{2+}$ ions from the second simulation set. It is found that the double ionized $Cu^{2+}$ from channel $Cu \rightarrow Cu^{2+}$ is located either near the cathode or in the outward region, as shown in Figure 7(a). The reason is due to high density of neutrals near the cathode or high electron energy in the outward region. Because electrons are accelerated toward the anode and gain energy, the electrons in the outward region tend to have enough energy to overcome the energy threshold of double ionization (about 28 eV). However, the competing channels of $Cu^+ \rightarrow Cu^{2+}$ also exist with a lower threshold energy (20.3 eV) and larger cross section. Therefore, $Cu^{2+}$ ions generated by this channel dominate in most area of the simulation domain, as shown in Figure 7(b).

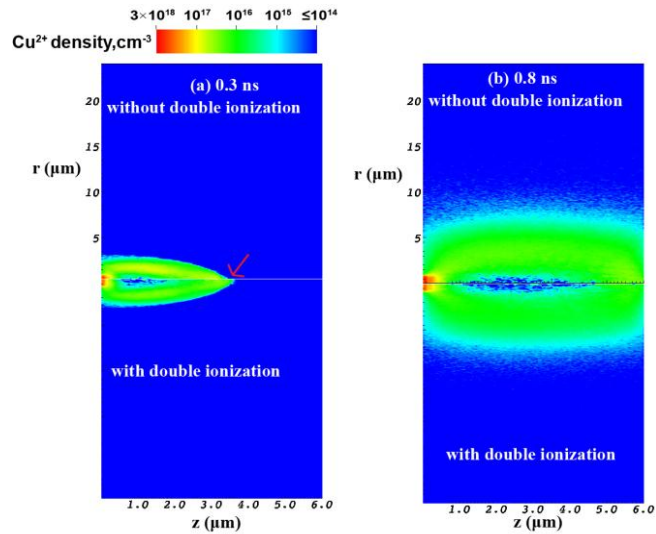

Figure 6. Number density of $Cu^{2+}$ ions at a time of (a) 0.3 ns, and (b) 0.8 ns respectively. The upper part is without double ionization from the first simulation set, and the lower part is with double ionization from the second set (mirror image along the symmetric axis *r*=0 for comparison). For both figures, the cathode is on the left, and the anode is on the right.

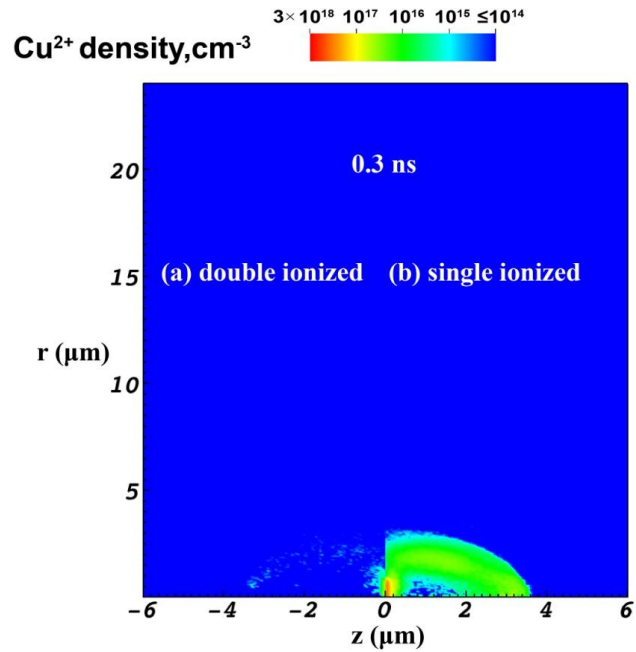

Figure 7. The spatial distribution of number density of differently labelled $Cu^{2+}$ ions from the second simulation set at a time of 0.3 ns (a) double ionized Cu -> $Cu^{2+}$, (b) single ionized $Cu^+$ -> $Cu^{2+}$. Figure 7(a) is a mirror image along the cathode for comparison.

## 3.3 Energy spectra of $Cu^{2+}$ ion

Figure 8 compares the energy spectra of $Cu^{2+}$ ions in the entire simulation domain for both sets at a time of (a) 0.3 ns and (b) 1.0 ns, respectively. At 0.3 ns, there are 71101 super-particles of $Cu^{2+}$ if double ionization is not considered, compared to that of 71342 super-particles of $Cu^{2+}$ if double ionization is considered. The most probable energy for $Cu^{2+}$ ions with double ionization is around 100 eV, and that without double ionization generally shifts to a higher energy around 130 eV. However, the shape of ion energy spectra at 1.0 ns almost coincide with each other for the two simulation sets, with most probable energy around 40 eV. The decrease of most probable energy of $Cu^{2+}$ ions during the expansion of cathode plasma toward anode is consistent with the simulation results in [20], in which the ion velocity along the symmetry axis can be accelerated to $2 \times 10^4$ m/s (corresponding to kinetic energy 133 eV) and then quickly falls.

The energy spectra of labelled $Cu^{2+}$ ions generated by double and single ionization of the second simulation set is shown in Figure 9. At 0.3 ns in Figure 9(a), it is found that the most probable energy of double ionized $Cu^{2+}$ ions is much lower than that of single ionized $Cu^{2+}$ ions. Because electron mass is much less than that of heavy particles, the electron-impact ionization process does not alter the velocity of heavy particle notably. Therefore, the double ionized $Cu^{2+}$ ions inherit velocity almost from that of slow Cu neutrals, while the single ionized $Cu^{2+}$ ions inherit velocity almost from that of fast $Cu^+$ ions. Neutral Cu particle cannot gain energy directly from the electric field. Therefore, those double ionized $Cu^{2+}$ ions have a low energy on average. After long time evolution to about 1 ns, the memory effect of initial velocity is not obvious anymore in the expansion of cathode plasma toward anode. Therefore, similar shape of energy spectra is observed for $Cu^{2+}$ ions generated from both single ionization and double ionization, as shown in Figure 9(b).

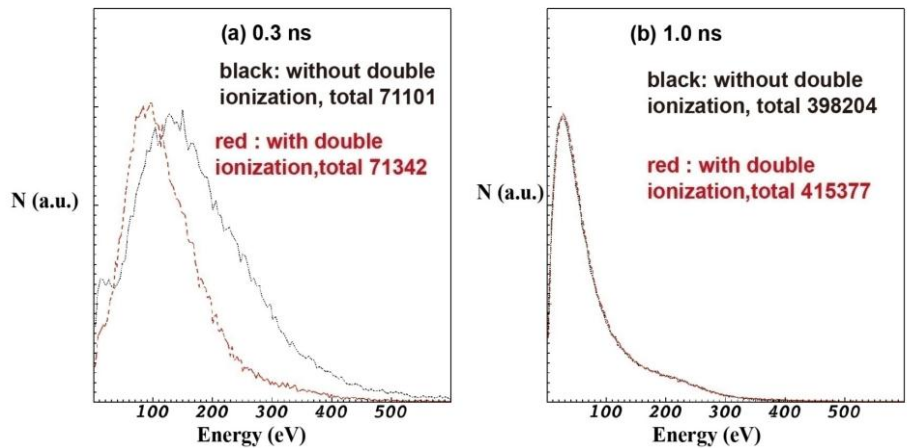

Figure 8. Energy spectra of $Cu^{2+}$ ions of both simulation sets at a time of (a) 0.3 ns and (b) 1.0 ns.

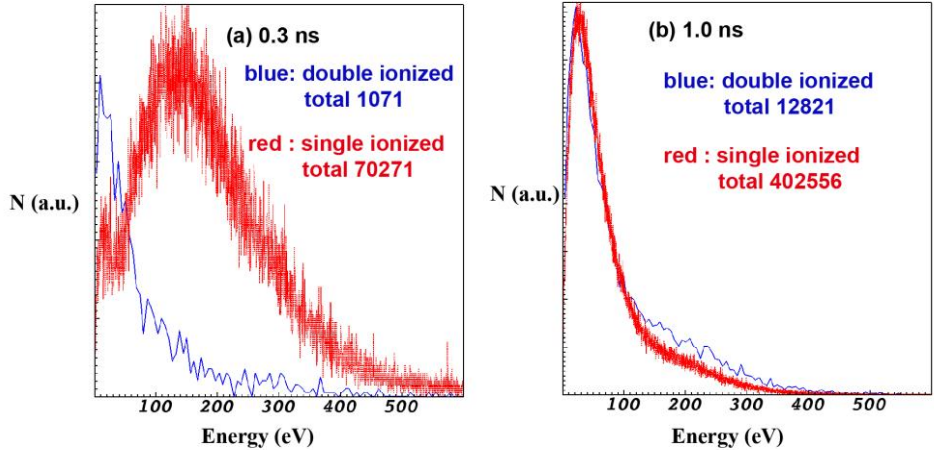

Figure 9. Energy spectra of labelled $Cu^{2+}$ ions generated by double and single ionization of the second simulation sets at a time of (a) 0.3 ns and (b) 1.0 ns.

## 4. Conclusions

In summary, a DSMC scheme for double ionization is proposed which conserves momentum, energy, charge, and mass during the collision. The DSMC scheme is then incorporated into a 2D3V PIC code to investigate the role of double ionization in the vacuum arc breakdown. Besides double ionization, the elastic collision between electron and neutral, step-wise ionization between electron and neutral, elastic collision between neutral and neutral, charge exchange between neutral and single charged ion, Coulomb collision are also included. A module for external circuit is coupled to the PIC simulation. The PIC-DSMC modelling starts from complete vacuum between two plane electrodes. An artificial tip with nominal field enhancement is assumed in the centre of the cathode, where field-emission electrons are injected into the simulation domain. A constant ratio of neutral flux to electron current density is used, but the calculated cathode erosion rate is time-dependent with value between 20~40 μg/C. Based on the simulation result of arc voltage and current, it can be divided into three stages as initiation, breakdown, and burn. Although the general discharge behaviour is not influenced in arc burning regime, the number density and energy spectra of $Cu^{2+}$ ions are altered in breakdown initiation with or without consideration of double ionization. This paper provides a quantitative research method to evaluate the role of multiply ionization in vacuum arcs.

**Acknowledgements**

This work is partially supported by NSFC Project Nos. 11875094 and 12005023, as well as the Foundation of President of China Academy of Engineering Physics Project No.YZJJLX2019013.

**Data availability statement**

All data that support the findings of this study are included within the article.